\documentclass[12pt]{article}

\usepackage{graphics, color}
\usepackage{graphicx}
\usepackage{amssymb}
\pdfoutput=1

\newcommand{\sect}[1]{\section{#1}\setcounter{equation}{0}}

\def\gsim{\, \rlap{$>$}{\lower 1.1ex\hbox{$\sim$}}\,}
\def\lsim{\, \rlap{$<$}{\lower 1.1ex\hbox{$\sim$}}\,}

\def\Op{{\mathcal{O}}}

 \newcommand{\be}{\begin{equation}}
\newcommand{\ee}{\end{equation}}
 \newcommand{\bal}{\begin{align}}
 \newcommand{\eal}{\end{align}}
\newcommand{\ben}{\begin{equation*}}
\newcommand{\een}{\end{equation*}}
\newcommand{\bea}{\begin{eqnarray}}
\newcommand{\eea}{\end{eqnarray}}
\newcommand{\bean}{\begin{eqnarray*}}
\newcommand{\eean}{\end{eqnarray*}}
\newcommand{\bes}{\begin{subequations}}
\newcommand{\ees}{\end{subequations}}

\def\ti{\tilde}

\usepackage{epsfig}

\newcommand{\comment}[1]{}

\begin{document}

\begin{titlepage}

\begin{flushright}
NSF-KITP-10-082
\end{flushright}
\bigskip
\bigskip\bigskip\bigskip
\centerline{\Large New stability results for Einstein scalar gravity}
\bigskip\bigskip\bigskip
\bigskip\bigskip\bigskip

\centerline{{\bf Thomas Faulkner}}
\medskip
\centerline{\em Kavli Institute for Theoretical Physics}
\centerline{\em University of California}
\centerline{\em Santa Barbara, CA 93106-4030}\bigskip
\medskip
\centerline{{\bf Gary T. Horowitz}}
\medskip
\centerline{and}
\medskip
\centerline{{\bf Matthew M. Roberts}}
\medskip
\centerline{\em Department of Physics}
\centerline{\em University of California}
\centerline{\em Santa Barbara, CA 93106-4030}\bigskip

\bigskip
\bigskip\bigskip

%ABSTRACT

\begin{abstract}
We consider asymptotically anti de Sitter gravity coupled to a scalar field with mass slightly above the Breitenlohner-Freedman bound. This theory admits  a large class of consistent boundary conditions characterized by an arbitrary function $W$. An important open question is to determine which $W$ admit stable ground states. It has previously been shown that the total energy is bounded from below if $W$ is bounded from below and the bulk scalar potential $V(\phi)$ admits a suitable superpotential. We extend this result and show that the energy remains bounded even in some cases where $W$ can become arbitrarily negative. As one application, this leads to the possibility that in gauge/gravity duality,  one can  add a double trace operator with negative coefficient to the dual field theory and still have a stable vacuum.

\end{abstract}
\end{titlepage}
\baselineskip = 17pt
\tableofcontents
\setcounter{footnote}{0}

\sect{Introduction}

Motivated by gauge/gravity duality, there has been considerable interest in the stability of asymptotically anti de Sitter (AdS) spacetimes. We investigate this for a class of theories involving gravity coupled to a real scalar field.   It is known that for standard (Dirichlet) boundary conditions, there is a positive energy theorem provided the scalar potential $V(\phi)$ can be derived from a ``superpotential" $P(\phi)$ \cite{Boucher:1984yx,Townsend:1984iu}.   When the scalar mass is slightly above the Breitenlohner-Freedman (BF) bound \cite{Breitenlohner:1982bm}, there is a large class of consistent boundary conditions characterized by an arbitrary function $W$. An important open question is to determine which $W$ admit stable ground states. In many cases, AdS itself is unstable, but there are static negative energy solitons which could be the minimum energy ground state.  In fact, these theories have been called ``designer gravity" \cite{Hertog:2004ns} since one can choose $W$ such that the theory admits static solitons with any prescribed energies. 

A conjectured answer to the question of  which $W$ have stable ground states was given in \cite{Hertog:2004ns} and will be reviewed in section 2. The following special case has been proven \cite{Amsel:2006uf,Amsel:2007im}: If $W$ is bounded from below (and the scalar potential $V(\phi)$ can be derived from a suitable superpotential), then the total energy of all solutions is bounded from below. In this paper we will first prove a stronger minimum energy theorem which allows $W$ to be unbounded from below. We then show that modulo a plausible technical assumption, this improved bound is equivalent to the original conjecture in \cite{Hertog:2004ns}.

This result  has a number of interesting consequences.
For example,  Ishibashi and Wald studied linearized stability of AdS  with general linear boundary conditions depending on a parameter $\kappa$ \cite{Ishibashi:2004wx}. They found that for masses just above the BF bound, there is a critical value of $\kappa$ such that AdS is  stable if $\kappa$ is larger than this critical value but unstable otherwise. We will see that for a large class of potentials, including the simplest case of a purely quadratic potential $V = \Lambda+ {1\over 2} m^2 \phi^2$, there is a stable ground state for all $\kappa$.

As a second example, in gauge/gravity duality, boundary conditions determined by $W$ correspond to adding to the dual field theory a new potential term involving $W(\Op)$, where $\Op$ is the operator dual to the bulk field $\phi$. \cite{Witten:2001ua,Berkooz:2002ug}. Our result shows that in some cases, one can add  potentials which are unbounded from below and still have a stable ground state\footnote{At least the sector representing the dual gravity modes has a stable ground state. It is possible that there are instabilities not captured by the bulk metric and scalar field.}. This is spontaneous symmetry breaking, where the stable vacuum is shifted but not destroyed.  More generally, our static solitons can be viewed as  representing RG flow on the sphere rather than the more familiar RG flow in Minkowski space. The gravitational dual in the latter case is often singular at small radius unless the scalar potential has a second extremum. In contrast, the finite volume of the sphere regulates the RG flow, and one obtains nonsingular solutions even when $V(\phi)$ has only one extremum.

The fact that modifying the boundary conditions can lower the energy of the gravitational  ground state has been seen even without a scalar field. Starting with AdS in Poincare coordinates, if one compactifies one spacelike direction (and allows antiperiodic boundary conditions for fermions) one has a negative energy solution known as the  AdS soliton. It has been conjectured that this is the minimum energy among all solutions with these boundary conditions \cite{Horowitz:1998ha}. 

Although most of the paper will be focussed on which boundary conditions $W$ lead to stable ground states,  in section 7 we will show that even with $W = 0$, the question of when there is a positive energy theorem can be very counterintuitive. We consider a simple potential $ V = -3 + {1\over 2} m^2 \phi^2 + \lambda \phi^4$ and show that for some $m$, the positive energy theorem holds only if $\lambda$ is {\it less} than a critical value. Increasing $\lambda$ will actually destabilize the ground state. 

This paper is organized as follows. In the next section we introduce the boundary condition function $W$, the idea of  designer gravity, and state the conjecture for when a stable ground state exists. We also review the earlier proof that the energy is bounded when $W$ is. In section 3 we generalize this proof to allow unbounded $W$. Sections 4 and 5 contain discussions of the needed superpotentials and their role in ``fake supergravity". In section 6 we discuss the global solitons and relate the minimum energy theorem in section 3 to the designer gravity conjecture. For simplicity, most of the paper is focussed on four spacetime dimensions and one value of $m^2$. Section 7 generalizes these results to higher dimensions and various masses. The final section contains further discussion. The appendix  provides more details on the generalization to various masses and dimensions.

\setcounter{equation}{0}
\sect{Review of designer gravity and previous minimum energy theorem}

Consider Einstein gravity  coupled to a neutral scalar field with potential $V(\phi)$,

\be
S ={1\over 2}\int d^4x \sqrt{-g}\left[R-(\nabla\phi)^2 -2V(\phi) \right].\label{action}
\ee
We will set the AdS radius equal to one,  and write the potential as $V(\phi)= -3+{1\over 2} m^2\phi^2+\ldots$. We will also assume $V$ is even. We wish to study asymptotically (global) $AdS_4$ solutions, where the metric approaches
\be
ds^2 = r^2(-dt^2 + d\Omega) + {dr^2\over r^2}
\ee
and $\phi\rightarrow 0$ at large radius. In four dimensions, the BF bound is $m^2 = -9/4$. We are interested in masses satisfying
\be\label{massrange}
m^2_{BF} < m^2 <m^2_{BF}+1.
\ee
Masses below the Breitenlohner-Freedman bound are unstable for any boundary conditions,  and for $m^2\ge m^2_{BF}+1$, we have to use Dirichlet boundary conditions \cite{Breitenlohner:1982bm}. However, when we are within the window (\ref{massrange}) we are free to impose very general boundary conditions.

For explicitness we will focus on the case $m^2=-2$ which lies in the range (\ref{massrange}), since this arises in supergravity and leads to simple asymptotic behavior for $\phi$. Our arguments apply to other masses satisfying (\ref{massrange}) and other spacetime dimensions as we discuss in section 7. For $m^2=-2$, the asymptotic behavior of $\phi$ is
\be\label{asymphi}
\phi = {\alpha\over r} + {\beta\over r^2}+\ldots
\ee
where $\alpha$ and $\beta$ can in general depend on the other coordinates. These solutions are normalizable for all $\alpha$ and $\beta$, but to have unique evolution, one must specify a relation between them. Familiar boundary conditions are $\alpha = 0$ (often called Dirichlet) or $\beta = 0$ (often called Neumann), but one can pick  any smooth function $W(\alpha)$ and set 
\be\label{bdycond}
\beta = {dW\over d\alpha}
\ee
%(Under gauge/gravity duality, this boundary condition corresponds to adding a new potential term involving W to the field theory action.) 
A key question is: for which $W$ does the theory have a stable, minimum energy ground state? A conjectured answer was given in \cite{Hertog:2004ns} which we now review.

The ground state need not be AdS, but it must be a static, finite energy solution, i.e., a soliton. One expects the minimum energy solution to be spherically symmetric, so
consider static, spherically symmetric solutions to (\ref{action}). We will construct these solutions in section 6. For now, it sufficies to note that for every choice of $\phi$ at the origin, one can integrate out and obtain an asymptotically AdS solution.  $\phi$ will satisfy (\ref{asymphi}) for some choice of $\alpha$ and $\beta$, which must now be constant due to the symmetries.  In other words, these solitons define a curve in the $\alpha,\beta$ plane. This curve depends on the potential $V(\phi)$, and in many cases one can represent it as a function $\beta_0(\alpha)$.\footnote{This is always true for small $\alpha$, but for large $\alpha$ the curve might double back and not yield a single valued function of $\alpha$. We will consider potentials for which this does not happen. Examples will be given in  section 6.} Define
\be
W_0(\alpha) = -\int_0^\alpha \beta_0 (\tilde \alpha) d\tilde \alpha
\ee
This is independent of boundary conditions. Given any boundary condition $W$, define
\be
{\cal V}(\alpha) = W (\alpha)+ W_0(\alpha)
\ee
Note that at an extremum of ${\cal V}$, $W' + W_0' =0$ and so there is an $\alpha$ with $\beta = \beta _0$. In other words, there is a soliton which satisfies the boundary condition. It was shown in \cite{Hertog:2004ns} that the total energy of the soliton is simply the value of ${\cal V}$ at the extremum. 
Conversely, given any function ${\cal V}(\alpha)$ (with ${\cal V}(0) = 0$), if one sets $W = {\cal V} - W_0$ then the theory with boundary condition given by this $W$ will have solitons at each extremum of ${\cal V}$, with masses given by the value of ${\cal V}$ at the extrema. It was for this reason that these theories were called ``designer gravity".

The following two conjectures were made in \cite{Hertog:2004ns}:

{\it Conjecture 1: The theory with boundary condition $\beta = dW/d\alpha$ has a stable ground state provided  ${\cal V} $ has a global minimum ${\cal V}_{min}$.  

Conjecture 2: The minimum energy solution is the spherical soliton associated with   ${\cal V}_{min}$. }

The solution with minimum energy is expected to be static and spherically symmetric. Since there is a static, spherical soliton with energy ${\cal V}_{min}$, conjecture  2 is very plausible. We will prove conjecture 1 in section 6 modulo a reasonable technical assumption.

 It was shown in \cite{Amsel:2006uf,Amsel:2007im} that if $W$ itself is bounded from below (and $V(\phi)$ can be derived from a suitable superpotential), then the energy is bounded below by the minimum of $W$. Conjecture 1 is stronger, since $W_0(\alpha)$ typically grows with $\alpha$. At the time  the conjecture was made, only one potential was studied in detail, and for that, $W_0 \sim |\alpha|$ for large $\alpha$. This  allows only a limited class of unbounded $W$ to still have a stable ground state. We will see that typically, $W_0 \sim |\alpha|^3$.
 
 We now briefly review the proof that the energy is bounded when $W$ is. In the next section we will show how to extend it to cases where $W$ is unbounded from below. 
One first introduces a ``superpotential" $P(\phi)$ satisfying
\be\label{Peq}
V(\phi) = 2\left({dP\over d\phi}\right)^2 - 3P^2
\ee
Near $\phi = 0$, the authors of \cite{Amsel:2006uf,Amsel:2007im} assumed\footnote{This was called $P_-$ in \cite{Amsel:2007im}. There is also a solution $P_+ = 1 + \phi^2/2 +  O(\phi^4)$, but it was shown in \cite{Amsel:2007im} that this superpotential does not lead to a minimum energy theorem when $\alpha \ne 0$.}
\be\label{Pminus}
P(\phi) = 1 + {1\over 4}\phi^2 + O(\phi^4)
\ee
Following Witten \cite{Witten:1981mf}, one now introduces a modified covariant derivative on a Dirac spinor $\Psi$
\be
\hat\nabla_\mu \Psi = \nabla_\mu\Psi + {1\over 2} P(\phi)\Gamma_\mu \Psi
\ee
and defines the Nester two-form \cite{Nester1981}
\be
B_{\mu\nu} = \bar\Psi\Gamma_{[\mu}\Gamma_{\nu}\Gamma_{\rho]}\hat\nabla^\rho \Psi  + h.c.
\ee
Given a spacelike surface $\Sigma$ with boundary $C$, let $\Psi$ be an asymptotically constant solution to Witten's equation:
\be
\Gamma^i\hat\nabla_i \Psi = 0
\ee 
where the index $i$ runs only over directions tangent to $\Sigma$. We further require that $-\bar\Psi \Gamma^\mu \Psi$ approach $\partial/\partial t$ asymptotically. Such solutions were shown to exist in \cite{Hertog:2005hm,Amsel:2007im}. The spinor charge 
\be
Q = \int_C{}^*B
\ee
can now be written as a manifestly positive volume integral over $\Sigma$, so $Q\ge 0$. The last step is to relate $Q$ to the total energy of the spacetime. When $\alpha\ne 0$, there are scalar contributions to the total energy which must be taken into account. Choosing $C$ to be a cut at constant $t$ for simplicity, it was shown in \cite{Amsel:2007im} that
\be
E = Q + \oint [W(\alpha) +\alpha\beta ]+ \lim_{r\rightarrow \infty}\oint\left[ {1\over 2} r\alpha^2  - 2r^3(P-1)\right]
\ee
where the $r$ dependence has been written explicitly and the integrals are over unit spheres at large $r$.
Plugging in the form of $P$ (\ref{Pminus}) and asymptotic behavior of $\phi$ (\ref{asymphi}) the divergent terms cancel and one obtains
\be
E\ge \oint W \ge 4\pi \ {\rm inf}\ W
\ee
Thus, the energy is bounded from below for all bounded $W$.

\setcounter{equation}{0}
\section{Generalized minimum energy theorem}

We now want to generalize the above theorem to allow $W$ which are unbounded from below. The idea is simply to note that the ansatz for the superpotential (\ref{Pminus}) is too restrictive. The equation  for $P(\phi)$ (\ref{Peq}) allows a one parameter family of solutions \cite{Papadimitriou:2006dr}.  In fact, for small $\phi$
\be\label{Ps}
P_s(\phi) = 1 + {1\over 4}\phi^2 - {s\over 6} |\phi|^3 + O(\phi^4)
\ee
is a solution for any constant $s$. While this is not analytic\footnote{Supersymmetry requires an analytic superpotential, but supersymmetry is playing no role in our analysis. In particular, even when $V(\phi)$ arises from a consistent truncation of supergravity, our boundary conditions will typically break supersymmetry.} at $\phi=0$, it is still $\mathcal{C}^2$ and so satisfies (\ref{Peq}) at the origin. Further, $P$ and $P'$ are continuous at the origin which is all that is required in constructing the spinor charge. If $P_s$ exists globally, we can repeat the above argument with this $P$, and obtain
\be\label{boundE}
 E\ge \oint [W + {s\over 3} |\alpha|^3]
 \ee
 So as long as $W(\alpha) + s |\alpha|^3/3$ is bounded from below, the total energy will be also. It is no longer necessary for $W$ itself to be bounded. A key  question is for what range of $s$ will  solutions  $P_s$ exist globally\footnote{There is also the question of showing that asymptotically constant solutions of Witten's equation exist with these superpotentials. We believe that the proof in \cite{Amsel:2007im,Hertog:2005hm} will carry over to this case, but have not checked this in detail.}? We will explore this in the next section. We will see that for most potentials $V(\phi)$, $P_s$ does exist globally up to a  critical value $s_c > 0$. Thus we get
 \be\label{newbound}
 E \ge 4\pi \ {\rm inf}\left[W(\alpha) + {1\over 3} s_c |\alpha|^3\right]
 \ee
 Even in cases where $s_c < 0$ (see section 7.1 for an example), (\ref{newbound}) provides a minimum energy theorem for boundary conditions $W$ such that the right hand side remains bounded from below.
 
 To relate this improved minimum energy theorem to the conjecture in section 2, we need to first show that the potentials $P_s$ exist, and then show that the function $W_0(\alpha)$ obtained from the solitons grows like $ {1\over 3} s_c |\alpha|^3 $ for large $\alpha$. Both will be established in the next few sections.

\setcounter{equation}{0}
\section{Existence of superpotentials}

The defining equation for a superpotential $V(\phi) =2P'^2 - 3P^2$ can always be solved perturbatively near $\phi=0$, but the question of global existence is harder. For definiteness, we will consider $\phi>0$ below. To define $P$ for $\phi<0$, we require $P$ be even. Since we want $P$ to approach (\ref{Ps}) for small $\phi$, we must choose  the positive square root and solve 
\be
P'(\phi)=\sqrt{\frac{3P^2}{2}+\frac{V(\phi)}{2}}.\label{PRootEq}
\ee
 It will prove useful to rewrite the differential equation in terms of $y=P'$ and $\phi$ using the fact that $P=\sqrt{2P'^2/3-V(\phi)/3}$:
\be
y'(\phi)=\frac{V'(\phi)}{4y}+\sqrt{\frac{3y^2}{2}-\frac{3V(\phi)}{4}}\label{yRootEq}
\ee
Since this is a first order differential equation, we know that solutions never cross except at singular points. Eq. (\ref{Ps}) tells us we are interested in solutions that vanish near $\phi=0$,
\be
y=\frac{\phi}{2}-\frac{s}{2}\phi^2+\ldots
\ee
Given how $s$ enters the minimum energy theorem, we now are interested in what the largest value $s_c$ of $s$ so that $P(\phi)$  exists for all $\phi$. Due to the nonlinear nature of (\ref{PRootEq}) and (\ref{yRootEq}), the exact value can only be found numerically. However we can make some general statements. $P$ will exist for all $\phi$ unless $P' = 0$. Since solutions to (\ref{yRootEq}) never cross, if a solution $P_1$ exists globally with some value $s=s_1$, then all solutions with $s < s_1$ exist globally as well. So we can algorithmically find $s_c$ by numerically solving (\ref{PRootEq}) or (\ref{yRootEq}), starting from $\phi=0$, and increasing $s$ until $P$ no longer exists globally because $P'(\phi)$ reaches zero. 

\begin{figure}[h!]
\begin{center}
\includegraphics[scale=.7]{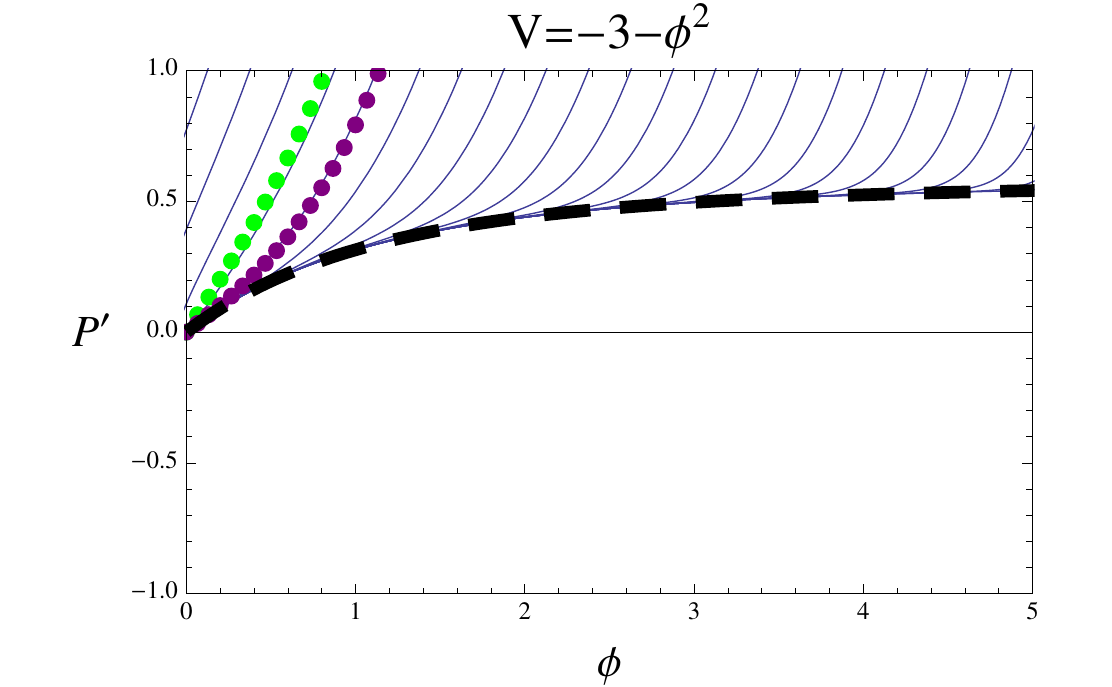}
\includegraphics[scale=.7]{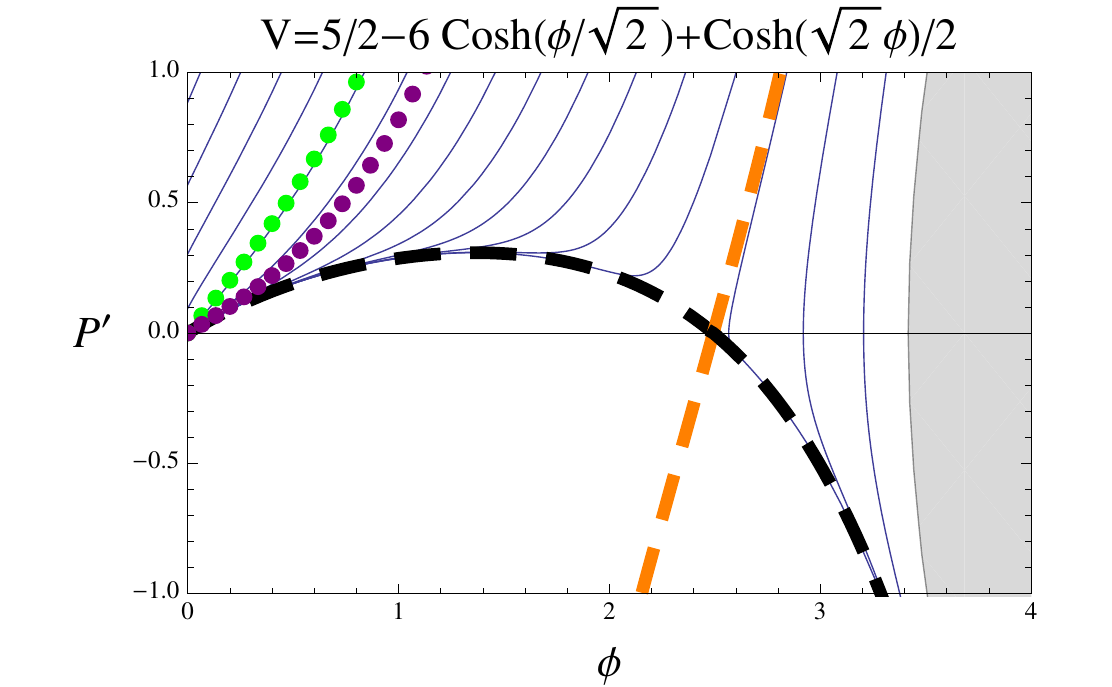}
\caption{
Solutions to (\ref{Peq}). Blue curves are generic solutions. The black dashed line is $P_c$, the solution with the largest value of $s_c$ which exists globally. The green and purple dotted lines are $P_\pm$ (see footnote 3). For $V=-3-\phi^2$, $s_c=0.52$. For $V=5/2-6\cosh(\phi/\sqrt{2})+\cosh(\sqrt{2}\phi)/2$, which comes from a consistent supergravity truncation \cite{Gauntlett:2009dn}, $s_c=.56$. The orange dashed curve in the right plot corresponds to $P^{IR}_+$  (recall $P^{IR}_-=P_c$.) The gray region in the right plot is where $2P'^2-V<0$ and therefore $P$ is not real.
 \label{fig:Fishbone} }
\end{center}
\end{figure}

 Generically, $P'$ vanishes at a value  $\phi_0$ where $V'\neq 0$. In this case, $3P^2/2+V(\phi)/2$ vanishes linearly, so $P'$ vanishes but $P''$ diverges.  One can solve for $P$ near $\phi_0$ to obtain
\be
P=\sqrt{-V(\phi_0)/3}\pm\sqrt{-2V'(\phi_0)/9}(\phi_0-\phi)^{3/2}+\ldots.
\ee
Note that when considering the contour $(\phi,P')$, these two branches meet smoothly, as can be seen in figure \ref{fig:Fishbone}. This solution does not exist for $\phi>\phi_0$ since the quantity inside the square root in (\ref{PRootEq}) becomes negative.

If $V$ has a second extremum $V(\phi_{IR}) =-V_0 < 0 ,~V'(\phi_{IR})=0$, it is possible to have $P'(\phi_{IR})=0$ and $P''(\phi_{IR})$ finite. Indeed if we expand (\ref{Peq}) around this extremum,
\be
P_\pm^{IR}=\sqrt{\frac{V_0}{3}}\left(1+\frac{\Delta^{IR}_\pm}{4}(\phi-\phi_{IR})^2+\ldots\right)\label{PIR}
\ee
where 
\be\Delta^{IR}_\pm=\frac{3}{2}\pm\sqrt{\frac{9}{4}+\frac{3V''(\phi_{IR})}{V_0}}.\ee
If $\phi_{IR}$ is a local minimum, $V''(\phi_{IR})>0$, then $\Delta_+^{IR}>0,~\Delta_-^{IR}<0.$ If we assume the global existence of the curves $P_\pm^{IR}$ (which appears to be generic in potentials with two extrema) then $P_-^{IR}=P_c$, because any $P$ with larger $s$  cannot intersect $P_\pm^{IR}$ and therefore $P'$ must vanish before $\phi_{IR}$, and therefore does not exist globally. For a clear example of this, see figure \ref{fig:Fishbone}.

For $s < s_c$, the curves run off to extremely large values of $P,~P'$. For these solutions $P^2,~P'^2\gg V$, and the bulk potential in (\ref{Peq}) becomes irrelevant. The solution for large $\phi$ then becomes
\be
P(\phi)=P_\infty e^{\sqrt{3/2}\phi}+\ldots.
\ee

If we have an unbounded potential which grows exponentially, there is also an asymptotic solution where we cancel the leading term in $V$. If we assume $V=-V_0 e^{b\phi}+\ldots$, then we find
\be
P=\sqrt{\frac{2V_0}{6-b^2}}e^{b\phi/2}+\ldots.
\ee
This implies that if $b>\sqrt{6}$, it is impossible to solve (\ref{Peq}) for large $\phi$ and no $P$ can exist globally. We now will show that any $P_s$ curve corresponds to a planar, boost invariant solution to the Einstein-scalar theory.
\setcounter{equation}{0}
\section{Fake supergravity}

Consider boost invariant planar solutions to our Einstein-scalar theory of the form\footnote{This is not the standard metric ansatz used in \cite{DeWolfe:1999cp,Freedman:2003ax} but we will use it to more easily compare to the global solitons in the next section.}
\be
ds^2=r^2(-dt^2+dx_i dx^i)+{dr^2\over g(r)}
\ee
and $\phi=\phi(r)$.  The constraint equations give an algebraic equation for $g$ which we can then eliminate from the equation for $\phi$ to obtain:

\be
g=\frac{2r^2V(\phi)}{r^2\phi'^2-6},\qquad \phi''+\frac{4\phi'}{r}-\frac{r\phi'^3}{2}+\frac{V_{,\phi}(\phi)}{V(\phi)}\left(\frac{3}{r^2}-\frac{\phi'^2}{2} \right)=0
\ee

It was noticed in \cite{DeWolfe:1999cp,Freedman:2003ax} that when the potential $V(\phi)$ can be derived from a superpotential $P$ as in (\ref{Peq}), the second order equation of motion can be reduced to a first order equation:
\be
\phi'(r)=-\frac{2P_{,\phi}(\phi(r))}{r P(\phi(r))}, \qquad g(r)=r^2P(\phi(r))^2.\label{fakeSUGRA}
\ee
This means that any superpotential generating our bulk potential $V$ corresponds to a special, boost invariant planar solution of the Einstein-scalar model. Since our boundary conditions require $\phi\rightarrow 0$ as $r\rightarrow \infty$, the behavior of $P_s$ near $\phi=0$ (\ref{Ps}) dictates the asymptotics of $\phi$ and $g$. Assuming $\alpha>0$ (general formulas are given in the appendix)

\be\label{asympsoln}
 \phi = \frac{\alpha}{r}-\frac{s\alpha^2}{r^2}+\ldots, \qquad g=r^2+\frac{\alpha^2}{2}-\frac{4 s\alpha^3}{3 r}+\ldots
 \ee
The first order equation (\ref{fakeSUGRA}) actually yields a one parameter family of solutions which can all be obtained by just rescaling $r \rightarrow cr, \ g\rightarrow c^2 g$. It is this scale invariance of the equations of motion which fixes the form of the curve $\beta(\alpha)= - s\alpha^2$  for these planar solutions.

In general, the energy density for planar solutions is  $E/V=M_0+\alpha\beta+W(\alpha)$ where $-M_0$ is the coefficient of the $1/r$ term in $g$.  (Note the close analogy to the energy of the global solitons \cite{Hertog:2004ns}.)  Using (\ref{asympsoln}) and $\beta(\alpha)= - s\alpha^2$ we see that the solution obtained from $P_s$ has energy
\be
{E\over V}=\frac{s\alpha^3}{3}+W.\label{fakeSUGRAEnergy}
\ee
Note that the solution corresponding to $P_c$ saturates an energy  bound analogous to (\ref{newbound}) for planar solutions. Solutions corresponding to $s<s_c$ would violate the bound, so they must have an interior singularity. This singularity must be such that there is no boundary condition on the spinor near the singularity which would make the boundary term in the spinor charge vanish. (Such boundary conditions are known for  black hole horizons \cite{Gibbons:1982jg}.) As shown in the last section, for $s< s_c$ one typically finds that $P$ blows up exponentially as $P=P_\infty e^{\sqrt{3/2}\phi}$. In this case, one can solve (\ref{fakeSUGRA}) and find
\be
\phi = -\sqrt{6}\log(r/r_0), \qquad g=\frac{P_\infty^2r_0^6}{r^4}
\ee
The scalar curvature diverges, $R=6 P_\infty^2r_0^6/r^6$. Ingoing null geodesics reach $r=0$ in finite $t$ and so it is a naked singularity. Therefore runaway $P$ fake supergravity solutions are not physical. 

One can ask if anything drastic occurs when $P'$ vanishes away from an extremum of the potential. However it is easy to calculate that $\phi$ and $g$ are smooth across $P'=0$. $\phi(r)$ simply has a local maximum there. One can in fact use the fake supergravity equations to interpret figure \ref{fig:Fishbone} as contour plots made from boost invariant solutions with different values of $s=-\beta/\alpha^2$. From this it is clear that such solutions will also have naked singularities, as the curves of $P'$ run off to negative infinity.

The only known case which does not have a singularity at small $r$ is the case where $V$ has a second extremum. In this case,  $P_c = P_{IR}^-$ and (\ref{PIR}) implies

\be
\phi=\phi_{IR}-\delta_{IR}r^{-\Delta^{IR}_-}+\ldots, \quad g=\frac{V_0}{3}r^2+\delta_{IR}^2\frac{V_0 \Delta^{IR}_-}{6}r^{2-2\Delta^{IR}_-}+\ldots
\ee
which  means that the small $r$ region of the solution is simply AdS with a shifted cosmological constant. Therefore the solution corresponding to $P_c$, the ground state for the system, is a domain wall.\footnote{In terms of gauge/gravity duality, this describes the RG flow of the original CFT perturbed by a relevant perturbation to a different IR CFT. }

In cases where the bulk potential does not have any other extrema, one may ask what the fake supergravity solution corresponding to $P_c$ corresponds to. This will in general depend on the form of the potential, but let us study as an example the case of a pure mass term, $V=-3+\frac{\Delta(\Delta-3)}{2}\phi^2$. From examining figure \ref{fig:Fishbone} it is clear $P_c'\approx {\rm const.}$, and expanding (\ref{Peq}) at large $\phi$ we find
\be
P_c=\frac{\sqrt{\Delta(3-\Delta)}}{\sqrt{6}}\phi+\Op(1/\phi),
\ee
which means from  (\ref{fakeSUGRA}) that the behavior near $r=0$ is
\be
\phi=2\sqrt{-\log(r)}+\ldots, \qquad g=\frac{2\Delta(\Delta-3)}{3}r^2\log(r)+\ldots
\ee
This is exactly the same behavior which arises in the zero temperature limit of the  ``large charge''  hairy black holes found in \cite{Horowitz:2009ij}.\footnote{This is because in that case the charged matter removed the electric field from the singular horizon and it is effectively neutral.} This is a null singularity, and since it is the zero temperature limit of regular AdS black holes,  it should be considered a physical state \cite{Gubser:2000nd}.

\setcounter{equation}{0}
\section{Scalar solitons}

Consider the most general spherically symmetric or plane symmetric static metric,

\be
ds^2=-f(r) dt^2+\frac{dr^2}{g(r)}+{r^2} d\Omega_k,\qquad \phi=\phi(r),
\ee
where $d\Omega_k$ is the metric on a $S^2$ with scalar curvature $2k$ (a unit $S^2$ has curvature 2.) The $k=0$ limit gives a metric on $\mathbb{R}^2$.
The equations of motion that follow from (\ref{action}) with this ansatz are
\bea
\frac{g'}{g}-\frac{f'}{f}+r\phi'^2&=&0, \nonumber\\
\frac{g'}{g}+\frac{f'}{f}+\frac{2 r V(\phi)}{g}+\frac{2}{r}-\frac{2 k}{rg}&=&0, \nonumber
\\
\phi''+\left(\frac{2}{r}+\frac{g'}{2g}+\frac{f'}{2f} \right)\phi'-\frac{V'(\phi)}{g}&=&0.\label{EOM}
\eea

 Recall that for our case of interest, $m^2=-2$ means $V(\phi)=-3-\phi^2+\Op(\phi^4)$. We can work out a large $r$ expansion to our equations of motion, yielding
\bea\label{asympexp}
\phi&=&\alpha/r+\beta/r^2+\ldots\nonumber\\
g&=&r^2+(k+\alpha^2/2)-M_0/r+\ldots
\nonumber\\
f&=&r^2+k-(M_0+4\alpha\beta/3)/r+\ldots
\eea
where $\alpha, \beta,$ and $M_0$ are undetermined constants. We wish to construct static scalar solitons (with $k\neq 0$). Smoothness at the origin implies $g= k+\ldots, \phi\approx \phi_0+\ldots$ Solving (\ref{EOM}) near the origin gives
\bea
\phi&=&\phi_0+\frac{V'(\phi_0)}{6k}r^2+\ldots\nonumber\\
g&=&k-\frac{V(\phi_0)}{3}r^2+\ldots\nonumber\\
f&=&f_0-f_0\frac{V(\phi_0)}{3k }r^2+\ldots\label{nearsoln}
\eea
Since we can always rescale $f$ (and $t$)  so that it asymptotes to $r^2+\ldots$ at the boundary, we are left with a one parameter family of static solitons found by varying $\phi_0$. From this we can read off $\alpha$ and $\beta$, and define $\beta_0(\alpha)$. Usually this requires numerically solving (\ref{EOM}), but in the limit of small $\phi$ and correspondingly small $\alpha, \beta$ we can solve it analytically. The perturbative solution is 
\be
g_{ab}=g_{ab}^{(AdS)}+\Op(\delta^2),~\phi=\delta\frac{\rm arctan(r)}{r}+\Op(\delta^3),
\ee
and therefore
\be
\beta_0=-\frac{2}{\pi}\alpha+\Op(\alpha^3).
\ee

The full nonlinear calculation was done in \cite{Hertog:2004ns} for a specific potential, $V=-2-\cosh(\sqrt{2}\phi)$. It was found that $\beta_0\rightarrow 0.71$ as $\alpha\rightarrow\infty.$ What we have found is that this is a special case of more general behavior, where as $\alpha\rightarrow\infty$ we asymptote to a scale-invariant relation $\beta_0\approx -s_w\alpha^2$ and we find in all cases that $s_w=s_c$. The supergravity truncation studied previously was a special case where $P_-=P_c$ and therefore $s_w=s_c=0$. Let us review this truncation and a simple generalization. We begin with the consistent truncation of $\mathcal{N}=8,~D=4$ supergravity to gravity and three neutral scalars \cite{Duff:1999gh},
\be
S=\frac{1}{2}\int d^4x\sqrt{-g}\left[ R-\sum_{i=1}^3(\nabla\varphi_i)^2+ \sum_{i=1}^3 2\cosh(\sqrt{2}\varphi_i)\right].
\ee
The previously studied truncation set $\varphi_2=\varphi_3=0$. For this truncation there is an analytic formula for $P_-=\cosh(\phi/\sqrt{2})$. We find numerically that no $P_s$ with $s>0$ exists globally for this potential, and so in this case $s_c=0$. However this is a special case of a more general truncation,
\be
\varphi_1=N\phi,~\varphi_2=b N \phi,~\varphi_3=c N \phi,~N=\frac{1}{\sqrt{1+b^2+c^2}},
\ee
which gives (\ref{action}) with
\be
V=-\cosh(\sqrt{2}\phi/N)-\cosh(b\sqrt{2}\phi/N)-\cosh(c\sqrt{2}\phi/N)=-3-\phi^2+\ldots\label{genSUGRApot}
\ee
We find that generically $s_c>0$, and the limit $b=c=0$ is a special case where $s_c\rightarrow 0$. Unfortunately, only in the case $b=c=0$ are we able to find an analytic formula for $P_c$.

\begin{figure}[h!]
\begin{center}
\includegraphics[scale=.7]{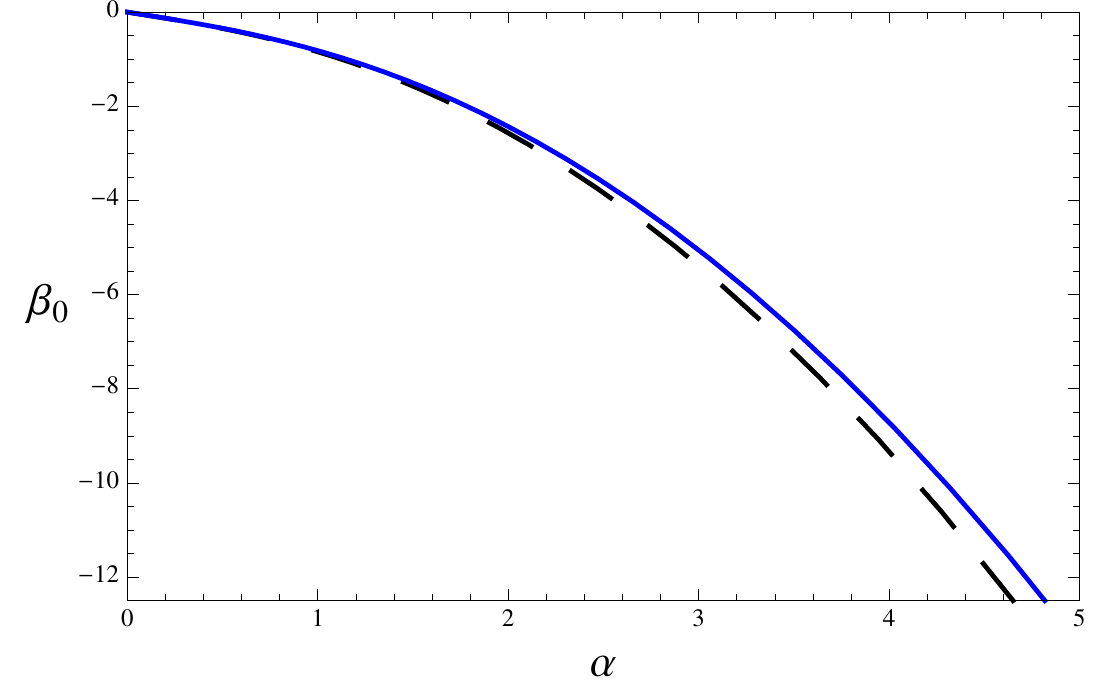}
\includegraphics[scale=.7]{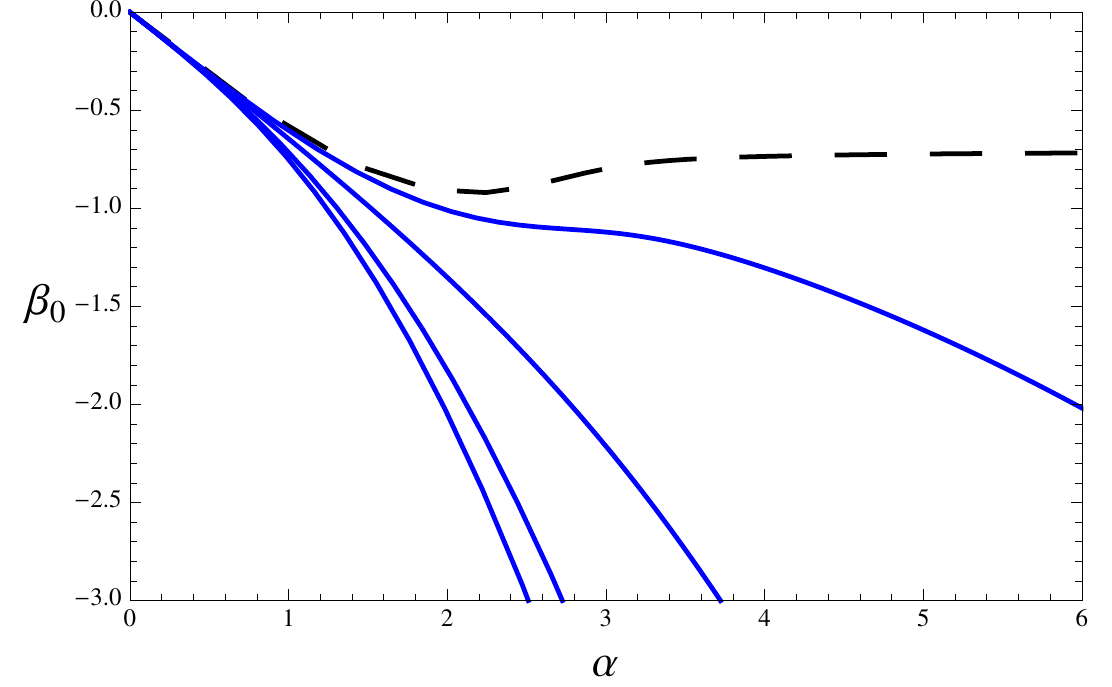}
\caption{
Plots of $\beta_0(\alpha)$ for various bulk potentials. All soliton curves have the same linear behavior at small $\alpha$ and have asymptotic scale invariance $\beta_0=-s_w\alpha^2$. The left figure is for  $V=-3-\phi^2,~s_w= 0.52$ (solid) and the potential used in \cite{Gauntlett:2009dn}\ $V=5/2-6\cosh(\phi/\sqrt{2})+\cosh(\sqrt{2}\phi)/2,~s_w=0.56$ (dashed.) The right figure is for (\ref{genSUGRApot}) with  for $b=c=1,~0.5,~0.25,~0.1$ (solid),  $b=c=0$ (dashed.) We find that as $b,c\rightarrow 0$, $s_w=s_c\rightarrow 0$. 
\label{fig:SolitonCurve} }
\end{center}
\end{figure}

In figure \ref{fig:SolitonCurve} we provide plots of $\beta_0$ for various potentials, and verify $s_w =s_c$. 

\subsection{Proving Conjecture 1}

We now relate the generalized minimum energy theorem in section 3 to  conjecture 1 in section 2.  We start with two preliminary results. Since $V$ is even, we can choose $\phi(0) > 0$ without loss of generality. We first show that $\phi(r)$ monotonically decreases to zero. Since $V(0)$ is a local maximum, $V' < 0$ for small positive $\phi$. If there is no other extremum, $V' < 0$ for all positive $\phi$. If there is a second extremum at $\phi_1$, then solutions which approach zero asymptotically must have $\phi(0)< \phi_1$. (This is because  (\ref{nearsoln}) shows that if $\phi(0) > \phi_1$, it will start increasing away from the origin  and  (\ref{EOM})  implies that $\phi$ cannot have a local maximum when $V' > 0$.) In either case, $V' < 0 $ over the region of interest. Thus from (\ref{nearsoln}),  $\phi(r)$ starts to decrease near the origin and from (\ref{EOM}), $\phi$ cannot have a local minimum or inflection point. Hence it must be monotonically decreasing. 

We next show that if $W_0(\alpha) = {1\over 3} s_w| \alpha|^3$ for any constant $s_w$, then $s_w \ge s_c$ and the coefficient of the $1/r$ term in $f$ vanishes. If we choose boundary conditions $W(\alpha)  = -W_0(\alpha)$, then all solitons satisfy our boundary conditions. Since we have a one parameter family of static solutions, and static solutions are extrema of the energy \cite{Sudarsky:1992ty}, all solitons have the same energy. Since the solitons approach AdS as $\phi(0) \rightarrow 0$, the total energy must be zero. Thus 
the new energy bound (\ref{newbound})  implies
\be 
W_0(\alpha) \ge {1\over 3}s_c|\alpha|^3
\ee
and hence $s_w \ge s_c$. Since the energy of the solitons is proportional to $M_0 + \alpha\beta + W$  we also learn
\be\label{useful}
M_0 + \alpha\beta -W_0 =0
\ee
The coefficient of the $1/r$ term in $f$ (from (\ref{asympexp})) is $-(M_0 + 4\alpha\beta/3) = -(\alpha\beta/3 + W_0)$. So if $W_0(\alpha) = {1\over 3} s_w |\alpha|^3$, $\beta = -s_w \alpha |\alpha|$ and this term vanishes.

We can now proceed to the main argument. We are interested in the behavior of $W_0(\alpha)$ for large $\alpha$. To extract it, we rescale
\be
  \alpha = c \ti \alpha, \quad r = c \rho, \quad \phi(c\rho) = \ti \phi(\rho),  \quad g(c\rho)  = c^2 \ti g(\rho), \quad f(c\rho)  = c^2 \ti f(\rho)
\ee
We are interested in the limit of large $c$ keeping $\ti \alpha$ fixed. 
Substituting this into the equations of motion (\ref{EOM}) we see that if $\phi(r), g(r), f(r)$ is a solution to the equations with $k=1$, then $\ti\phi(\rho), \ \ti g(\rho), \  \ti f(\rho)$ satisfies the equations with $k$ replaced by $\ti k = 1/c^2$. So in the limit of large $c$, we expect to approach a solution of the planar equations with $k = 0$. We now assume this is the case:

{\it Technical assumption: The limit $\ti k \rightarrow 0$ is not singular, i.e., we can  expand $\ti \phi = \ti \phi^{(0)} + \ti  k \ti \phi^{(1)} + \ldots $, and similarly for $\ti g$ and $\ti f$.}

For nonzero $\ti k$, the soliton solution depends on two scales, $\ti k$ and $\ti \alpha$. In the limit $\ti k \rightarrow 0$, $\ti \phi^{(0)}$ depends only on $\ti \alpha$. This means that in this limit, $\ti W_0$ must take the scale invariant form $\ti W_0 = {1\over 3} s_w |\ti \alpha|^3$ for some constant $s_w$.

 It follows that $\ti \phi^{(0)}$  must be a boost invariant solution of the planar equations and hence determined by the superpotential as in section 5. This is because the general translationally invariant planar solution depends on two constants, which appear in the $1/\rho$ expansion of the field equations at large $\rho$. But we showed above that the free constant in $f$ vanishes whenever $W_0$ is cubic. Thus $\ti f$ approaches $\rho^2$ in the limit.

To complete the argument, we now use the fact that boost invariant planar solutions are given in terms of the superpotential $P_s$. But if $s_w > s_c$, the superpotential does not exist globally.  $P'$ changes sign at some $\phi_{max}$ which means that $\phi(r)$ has a local maximum. But the global solitons cannot have a local maximum. Furthermore, the corrections between the global solitons and planar solutions do not show up until $\rho \approx 1/c$ whereas the local maximum occurs at an $\rho$ independent of $c$. This contradiction shows that $s_w = s_c$ and hence the generalized minimum energy theorem is equivalent to conjecture 1.

Proving conjecture 2 will require a different approach. One problem is that in the inequality for the spinor charge, equality holds only when there exist spinors satisfying $\hat\nabla_i \Psi = 0$. This might be satisfied for the planar domain walls, but cannot hold for the spherical soliton.

%%%%%%%%%%%%%%%%%
\setcounter{equation}{0}
\sect{Generalization to different masses and spacetime dimensions}

We have focussed on asymptotically $AdS_4$ solutions with $m^2 = -2$ for definiteness and simplicity. In this section, we show how the argument that there is a stable ground state for some unbounded $W$ extends to other masses and spacetime dimensions. Consider the following action in $d+1$ dimensions
\be
S ={1\over 2}\int d^{d+1}x \sqrt{-g}\left[R-(\nabla\phi)^2 -2V(\phi) \right].\label{actiond}
\ee
with potential  $V(\phi)= -{1\over 2} d(d-1)+{1\over 2} m^2\phi^2+\ldots$ where the mass again satisfies (\ref{massrange}) with $m^2_{BF} = -d^2/4$.

The asymptotic behavior of $\phi$ is
\be\label{asymphid}
\phi = {\alpha\over r^{\Delta_-}} + {\beta\over r^{\Delta_+}}
\ee
where
\be\label{deltadefd}
\Delta_\pm = {d\over 2} \pm \sqrt{{d^2\over 4} + m^2}
\ee
Boundary conditions are again described by an arbitrary function $W(\alpha)$ with $\beta = W'(\alpha)$. 

To establish a minimum energy theorem, one  introduces a superpotential $P(\phi)$ satisfying
\be\label{Peqd}
V(\phi) = (d-1)\left({dP\over d\phi}\right)^2 - d P^2
\ee
Near $\phi = 0$, 
\be\label{Pminusd}P_s(\phi)=\sqrt{\frac{d-1}{2}}+\frac{\Delta_-}{2\sqrt{2(d-1)}}\phi^2-s\frac{\Delta_-(\Delta_+-\Delta_-)}{d\sqrt{2(d-1)}}|\phi|^{d/\Delta_-}+\ldots
\ee
for any constant $s$. Note that since $\Delta_-<d/2$, $P_s$ is $\mathcal{C}^2$. When $P_s$ exists globally, one can use it to prove a generalized minimum energy theorem, 
\be
 E\ge \oint [(\Delta_+ - \Delta_-) W +\frac{s_c\Delta_-(\Delta_+-\Delta_-)}{d}|\alpha|^{d/\Delta_-}]
 \ee
(The details of the proof are given in the appendix.) So as long as the integrand  is bounded from below, the total energy will be also. As before,  the key remaining question is for what range of $s$ will  solutions  $P_s$ exist globally? One can again show  that for most potentials $V(\phi)$, $P_s$ does exist globally up to a  critical value $s_c > 0$. Similarly we find that for static global solitons, assuming a well behaved large $\alpha$ limit, 
\be\label{newboundd}
W_0=\frac{s_c \Delta_-}{d}|\alpha|^{d/\Delta_-}+...
\ee
then the large $\alpha$ limit turns global solitons into the boost invariant fake supergravity solution corresponding to $P_c$. Therefore $W+W_0$ is bounded when $W+\frac{s_c \Delta_-}{d}|\alpha|^{d/\Delta_-}$ is bounded and our minimum energy theorem proves conjecture 1.

 \subsection{A surprising minimum energy theorem}
 
We now discuss a curious feature which occurs when the dimension $\Delta_-$  of the operator $\Op$ dual to $\phi$ is close to the unitarity bound.
This feature is only present for $d \leq 3$. 
 For example in $d=3$, where the unitarity bound is $\Delta_- > 1/2$, we will see that
 the nature of the bulk potential required for a minimum energy theorem
 changes drastically (as a function of $\Delta_-$) when $\Delta_- = 3/4$. Interestingly 
 this is exactly when the multitrace
 operator $\mathcal{O}^4$ in the dual theory becomes relevant.  The minimum
 energy theorem we refer to in this section is for boundary conditions $\beta=0$ and thus
 relates to the global existence of $P_-$ or equivalently to the requirement that $s_c\ge 0$.\footnote{Recall that a positive energy theorem
 can exist even if $P_-$ does not exist, however it requires  special boundary conditions $W\neq 0$ which regulate the effective potential. } 
 
For example consider an arbitrary scalar potential with an expansion
about $\phi=0$,
 \be
 \label{eq:v}
 V(\phi) = -3 + {1\over 2} m^2 \phi^2 + \frac{\lambda}{4} \phi^4 + \ldots
 \ee
Normally, the superpotential $P_-$ exists (and hence a positive energy theorem can be proved) if 
$\lambda$ is larger than a critical value $\lambda_c $ (see, e.g., \cite{Hertog:2006rr}). However, when $\Delta_-$ is smaller than $3/4$, the situation is reversed. A positive energy theorem exists only if $\lambda$ is smaller than some critical value! 
 
 To see this, we solve $V = 2P'^2 - 3P^2$ for the superpotential $P_-$ for small $\phi$. Since we are setting $W=0$, we take $s=0$. Expanding to $O(\phi^4)$, the solution is 
 \be
 P_- = 1 + {\Delta_-\over 4}\phi^2 + {4\lambda + 3\Delta_-^2\over 32(4\Delta_- -3)}\phi^4+\ldots
 \ee
 So for $\Delta_- > 3/4$, increasing $\lambda$ increases the coefficient of the $\phi^4$ term and makes it less likely that $P'$ will vanish. However, for $\Delta_- < 3/4$, increasing $\lambda$ {\it decreases} the coefficient of the $\phi^4$ term making it more likely that $P'$ will vanish. From (\ref{deltadefd}) this will be the case for $m^2$ close to $m^2_{BF} + 1$.  
 
 To illustrate this, in Fig. 3 we show the curves of $P'$ vs. $\phi$ for the case $\Delta = 3/5 < 3/4$ 
with two different values of $\lambda$ and no higher order terms in (\ref{eq:v}) . Since we have set $s=0$, we are interested in the $P_-$ curve indicated in purple (dotted.) The blue curves correspond to nonzero values of $s$. In the first case, $\lambda > 0$ and the $P_-$ curve crosses the axis before the second extremum, showing that there is no minimum energy theorem.
  In the second figure, $\lambda < 0 $ and there is a minimum energy theorem.
 
 \begin{figure}[h!]
\begin{center}
\includegraphics[scale=.9]{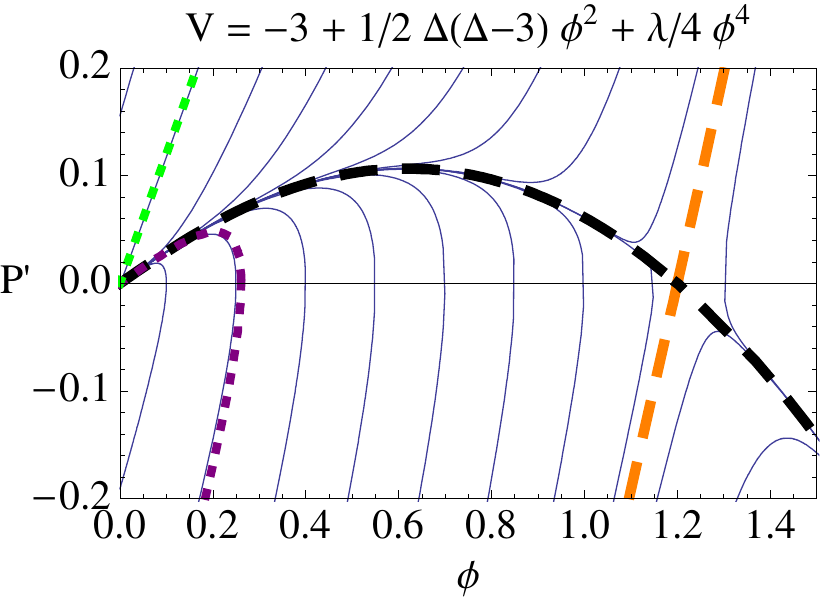}
\includegraphics[scale=.9]{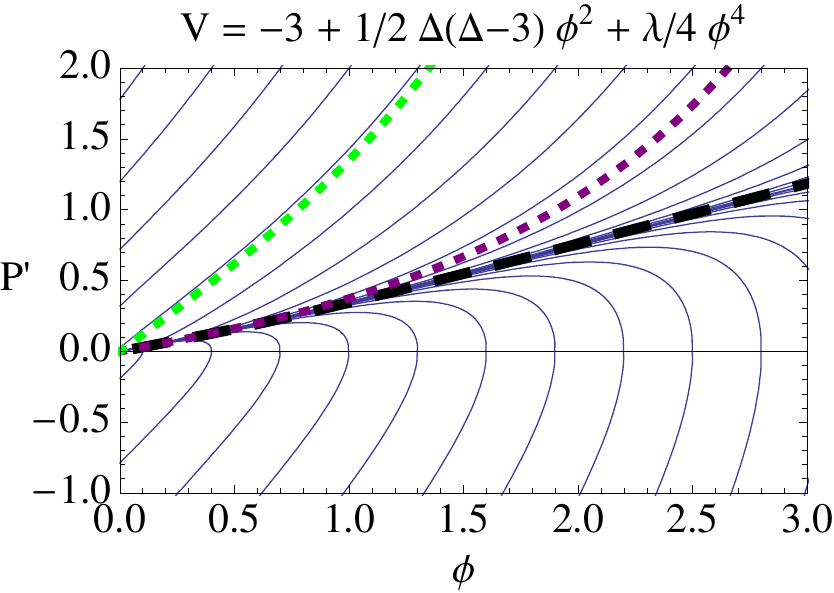}
\caption{
Plot of $P'$ analogous to Fig. 1 for a $\phi^4$ potential
with $\Delta=3/5$ and two values of $\lambda$. (\emph{left}) The $P_-$ curve crosses the axis before the second extremum, showing that there is no minimum energy theorem for Neumann
boundary conditions.  (\emph{right}) Here the $P_-$ exists globally and 
there is a minimum energy theorem
for Neumann boundary conditions. 
Because in this case $\lambda$ is negative there is no second extremum
in this potential however the critical curve $P_c$ still exists.
\label{fig:phasebad} }
\end{center}
\end{figure} 

The exact critical value $\lambda$ for which the minimum energy theorem is lost
depends on details of the potential and on $\Delta_-$. 
However we can make the condition on $\lambda$ more precise by studying the 
behavior of $s_c$ as
a function of $\Delta_-$ close to $\Delta_- = 3/4$. 
%As mentioned earlier, if
%$s_c$ is negative then a positive energy theorem does not exist with boundary condition $\beta=0$. 
The generalized superpotential is,
  \be
  P_{c} = 1 + {\Delta_-\over 4}\phi^2 - s_c \frac{ \Delta_- (3 - 2 \Delta_-)}{6}| \phi|^{3/\Delta_-} +   {4\lambda + 3\Delta_-^2\over 32(4\Delta_- -3)}\phi^4+\ldots
  \ee
 
 Since $P_c$ must be globally defined, the pole at $\Delta_-=3/4$ in the $\phi^4$ term must be canceled by the $|\phi|^{3/\Delta_-} $ term when
 we take the limit $\Delta_- \rightarrow 3/4$. This will only occur if $s_c$ (more
 generally any $P_s$ must have this property) also has a pole,
 
 \begin{equation}
 \label{eq:pole}
 s_c  \sim - \frac{ 2[ \lambda + (3/4)^3] }{3 (4 \Delta_- -3)} + \ldots
 \end{equation} 
  
  Thus for $\Delta_-$ close to $3/4$ the critical value of $\lambda$ approaches
  $\lambda_c = -(3/4)^3$ such that the existence of a positive energy
  theorem boils down to the requirement that
   $ (\lambda+ (3/4)^3) \times (\Delta_- - 3/4) > 0$. We illustrate
  this in Fig.~4 where $s_c$ is plotted as a function of $\Delta_-$ for two
  different values of $\lambda$.

 \begin{figure}[h!]
\begin{center}
\includegraphics[scale=1.2]{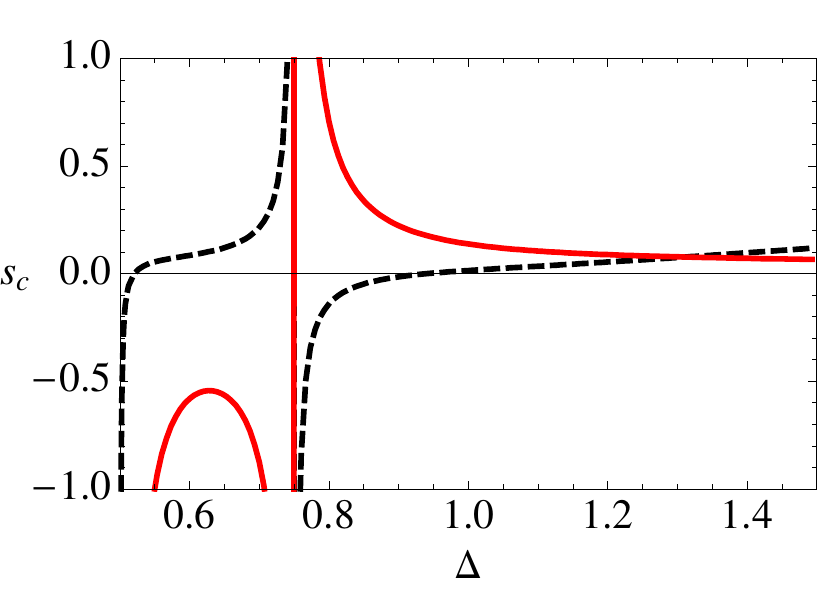}
\caption{
Plots of $s_c$ as a function
of $\Delta_-$ for two different choices of $\lambda$. The solid red
curve is for $\lambda = 1$ and the dashed black curve is
for $\lambda = -1/2$. 
Recall that
the condition for a minimum energy theorem with $\beta = 0$ boundary condition is $s_c \ge 0$. Close to $\Delta_- = 3/4$ the behavior of $s_c$ is determined
by the pole in (\ref{eq:pole}). The sign of the coefficient of the pole is different
for the two cases presented here.
\label{fig:sdelta} }
\end{center}
\end{figure}

It is likely that the existence/non-existence of a minimum energy theorem
is at the heart of the behavior observed in Section II B of \cite{Iqbal:2010eh} where
the asymptotic behavior of their $B(A)$ curves (analogous to the $\beta(\alpha)$ soliton
curves in Section B) changed dramatically as they tuned across
the point $\Delta_- = 3/4$. 

 Note for higher space-time dimensions, say $d=4$, there are no multi-trace operators of the form $\mathcal{O}^{2n}$
 for $n>1$ that can become relevant while obeying the unitarity bound. (The operator $\Op^2$ is always relevant when $\Delta_-<d/2$.) Thus
 the above feature is not present in higher dimensions. While in $d=2$
 since the unitarity bound is $\Delta_-  \geq 0$ there are potentially an infinite set
 of multi trace operators $\mathcal{O}^{2n}$ which can be relevant as we move
 towards $\Delta_- = 0$. There will be
 similar behavior as that discussed above; each time a new multi trace
 operator becomes relevant $s_c$ switches sign (via a pole) and the
 term in the expansion of the bulk potential $\lambda_{2n} \phi^{2n}$ determines the behavior of $s_c$  close to this pole.

%%%%%%%%%%%%%
\setcounter{equation}{0}
 \section{Discussion}
  We have considered gravity coupled to a scalar with mass close to the BF bound and proved a generalized minimum energy theorem (section 3). This shows that the total energy remains bounded from below even in cases when the function   $W(\alpha)$ determining the boundary condition becomes arbitrarily negative. Modulo a plausible assumption about a planar limit, we showed in section 6 that our result establishes  a conjecture  made in \cite{Hertog:2004ns}.
 
 We also saw in section 6 that for small $\phi$, the soliton equations can be solved exactly, and yield $\beta_0(\alpha) = -2\alpha/\pi$. This is equivalent to a result by Ishibashi and Wald \cite{Ishibashi:2004wx} who studied linearized stability of AdS with linear boundary conditions $\beta = \kappa \alpha$. They found that AdS is stable for $\kappa > -2/\pi$ and unstable for $\kappa < -2/\pi$. (The static soliton lies on the boundary of these two regiemes.) We  now understand the nonperturbative situation with boundary condition $\beta = \kappa \alpha$. If the potential $V(\phi)$ is like the ones considered in Fig. 2, then there is a stable ground state for all $\kappa$. If the potential is like the ones on the right side on Fig. 2 with $b$ close to $\sqrt 2$, then for some $\kappa$, AdS is stable, but there is still a lower energy ground state!
 
 We now discuss a couple of easy generalizations of the minimum energy results above. First, one can choose the solution to Witten's equation such that asymptotically, $-\bar\Psi \Gamma^\mu \Psi$ approaches the timelike Killing field $(\partial/\partial t) + \Omega \partial/\partial \varphi$ for any $|\Omega| < 1$. Repeating the argument leads to inequalities analogous to  (\ref{newbound}) and (\ref{newboundd}) with $E$ replaced by 
 $E - |J|$ where $J$ is the angular momentum associated with the rotation $\varphi$.
 
 As a second generalization, we have discussed gravity coupled to a real scalar, but the extension to a charged scalar is straight-forward. The spinor charge remains positive if we couple gravity to any matter satisfying the dominant energy condition ($T_{\mu\nu} t^\mu \hat t^\nu \ge 0$ for all timelike $t^\mu,\hat t^\nu$). The action for a charged scalar satisfies this except for the possibility of a potential term which is only a function of $|\phi|$. By modifying the covariant derivative on the spinor as we have done here, one can take care of the potential and obtain a lower bound on the energy.

A third possible generalization is to saturating the Breitenlohner-Freedman bound. In \cite{Amsel:2006uf}, the authors found that in the limit where  the bound is saturated, an additional divergent term appears which ruins the minimum energy theorem when using $P_-$. However, the limit of (\ref{Pminusd}) will give an additional term of order $\phi^2\log(\phi)$ in $P$, which can perhaps be used to cancel the divergence.
 
 We now turn to implications of our result for the dual field theory. One of the main corollaries of the energy theorem that we have discussed in this paper
 is the possibility of deforming the theory with boundary conditions which destabilize AdS yet still
 have a stable ground state\footnote{There is a large literature on the stability of multitrace deformations from the standpoint of the dual field theory. See, e.g., \cite{Vecchi:2010dd} and references therein.}. For example, deforming by a (relevant) double trace operator 
 $ \frac{\kappa}{2} \mathcal{O}^2$  
with negative coefficient $\kappa < 0$ (corresponding to linear boundary conditions $\beta=\kappa\alpha$), will yield a symmetry breaking ground state.
Since the deformation does
 not explicitly break the $Z_2$ symmetry $\mathcal{O} \rightarrow - \mathcal{O}$ the
 breaking is spontaneous. The operator $\mathcal{O}$ gets a vev in the new ground state
 which scales like $(-\kappa)^{\Delta_-/(d-2\Delta_-)}$.
 This is fairly natural thing to do in a CFT, a classic textbook example being the Wilson Fisher
 fixed point. More generally our energy bounds (\ref{newbound}) and (\ref{newboundd})
 allow for the inclusion of all \emph{relevant} 
 multi-trace operators with arbitrary coefficient. 
 This includes for $d=3$ and $\Delta_-<3/4$ the operator
 $\mathcal{O}^4$, thus providing an example of a
 set of deformations analogous to that of a tricritical point. 
 
 More generally, one can apply this analysis to a complex scalar field in the bulk dual to a complex operator $\mathcal{O}$. A double trace deformation with negative coefficient now spontaneously breaks a $U(1)$ symmetry. The existing constructions of holographic superconductors  \cite{Hartnoll:2008vx} achieve this using nonzero chemical potential and a charged black hole in the bulk. We now have a new mechanism for obtaining a charged condensate at low temperature without turning on a chemical potential. The critical temperature is proportional to $(-\kappa)^{1/(d-2\Delta_-)}$. Various properties of these models and other condensed matter applications of multitrace deformations will be discussed in a forthcoming paper \cite{us}.

 Another natural question relates to  the nature of the ground state that results from this deformation. This depends on the details of the bulk potential and cannot be answered
 in generality. If the potential has a global minimum then the deformation should describe
 a flow to another CFT. This is not the case for unbounded potentials  and this state
 may represent something more interesting on the field theory side.
In particular, it would be interesting to find examples where the end state
 was confining. It seems likely this would require a first order transition and
 thus it is not clear how it will fit into our story.

 Having said all this it is important to emphasize that our results apply to bulk theories truncated to
 purely gravity plus scalar. It is possible that in the fully string theoretic realizations
 these boundary conditions lead to an instability outside of this truncation. An example
 of such an instability is to the fragmentation of branes from behind the Poincare horizon.
 This would correspond in the field theory side to matrix eigenvalues being forced
 out on to the coulomb branch. Since many  CFTs with gravity duals
 have flat directions this remains a real possibility.
 The question of stability on the gravity side becomes
 that of studying the force on a probe brane. We cannot study this within our
 current setup and leave this question to the future. 
 If there, this instability would be analogous  to various instabilities that arise
 when turning on a non-zero chemical potential in gravity duals (see, e.g., \cite{Yamada:2008em}).
 
\vskip 1cm
\centerline{\bf Acknowledgements}
\vskip .5 cm
It is a pleasure to thank D. Marolf, H. Liu, N. Iqbal, and E. Silverstein for discussions. This work was supported in part by the US National Science Foundation under Grant No.~PHY08-55415, Grant No. PHY05-51164
and the UCSB Physics Department.

\begin{appendix}
\setcounter{equation}{0}
\section{Proof of the minimum energy theorem}
In this appendix, we give more details of the minimum energy theorem in sec. 7 for general mass and spacetime dimension.
Consider a general Einstein-scalar system, with action
\be
S=\frac{1}{2}\int d^{d+1}x\sqrt{-g}\left[R-(\nabla\phi)^2-2V(\phi) \right]\label{gendimaction}
\ee
where $V=-\frac{d(d-1)}{2}+\frac{m^2}{2}\phi^2+\Op(\phi^4)$\footnote{Note that here we have chosen coordinates where the AdS radius $L=1$.}, and we are in the window 
\be
 m^2_{BF}<m^2<m_{BF}^2+1, ~\qquad m_{BF}^2=-d^2/4.
 \ee
 The asymptotic behavior of $\phi$ is
\be\label{asymphid}
\phi = {\alpha\over r^{\Delta_-}} + {\beta\over r^{\Delta_+}}
\ee
where
\be\label{deltadefda}
\Delta_\pm = {d\over 2} \pm \sqrt{{d^2\over 4} + m^2}
\ee
When the mass is in the above range,  $\frac{d-2}{2}<\Delta_-<\frac{d}{2}.$
Boundary conditions are again described by an arbitrary function $W(\alpha)$ with $\beta = W'(\alpha)$. To establish a minimum energy theorem, one  introduces a superpotential $P(\phi)$ satisfying
\be\label{Peqd}
V(\phi) = (d-1)\left({dP\over d\phi}\right)^2 - d P^2
\ee
Near $\phi = 0$,\footnote{There is also a branch of solutions of the form $P_+=\sqrt{\frac{d-1}{2}}+\frac{\Delta_+}{2\sqrt{2(d-1)}}\phi^2+\ldots$, however this does not give us a finite energy bound \cite{Amsel:2007im}, and only $s=0$ is 
 allowed.} 
\be
P_s(\phi)=\sqrt{\frac{d-1}{2}}+\frac{\Delta_-}{2\sqrt{2(d-1)}}\phi^2-s\frac{\Delta_-(d-2\Delta_-)}{d\sqrt{2(d-1)}}|\phi|^{d/\Delta_-}+\ldots\label{genPs}
\ee
for any constant $s$. We call the largest value of $s$ for which $P$ exists globally $s_c$.
Define a modified covariant derivative on a Dirac spinor $\Psi$
\be
\hat\nabla_\mu \Psi = \nabla_\mu\Psi + {1\over \sqrt{2(d-1)}} P(\phi)\Gamma_\mu \Psi
\ee
and let
\be
B_{\mu\nu} = \bar\Psi\Gamma_{[\mu}\Gamma_{\nu}\Gamma_{\rho]}\hat\nabla^\rho \Psi  + h.c.
\ee
Given a spacelike surface $\Sigma$ with boundary $C$, let $\Psi$ be an asymptotically constant solution to Witten's equation:
\be
\Gamma^i\hat\nabla_i \Psi = 0
\ee 
We further require that $-\bar\Psi \Gamma^\mu \Psi$ approach $\partial/\partial t$ asymptotically. Such solutions were shown to exist in \cite{Hertog:2005hm,Amsel:2007im}. The spinor charge 
\be
Q = \int_C{}^*B
\ee
can now be written as a manifestly positive volume integral over $\Sigma$, so $Q\ge 0$. The last step is to relate $Q$ to the total energy of the spacetime. Choosing $C$ to be a cut at constant $t$ for simplicity, it was shown in \cite{Amsel:2007im} that
\be
E = Q + \oint [(\Delta_+ - \Delta_-) W(\alpha) +\Delta_- \alpha\beta ]+ \lim_{r\rightarrow \infty}\oint r^{d-2\Delta_-} {\alpha^2\Delta_- \over 2}  - r^d\sqrt{2(d-1)} \left (P-\sqrt{{d-1\over 2}}\right )
\ee
where the integrals are over unit spheres at large $r$.
Plugging in the form of $P$ (\ref{genPs}) and asymptotic behavior of $\phi$ (\ref{asymphid}) the divergent terms cancel and one obtains 
\be\label{boundEd}
 E\ge \oint [(\Delta_+ - \Delta_-) W +\frac{s_c\Delta_-(\Delta_+-\Delta_-)}{d}|\alpha|^{d/\Delta_-}]
 \ee
 So as long as the integrand  is bounded from below, the total energy will be also. As before,  the key remaining question is for what range of $s$ will  solutions  $P_s$ exist globally? One can again show  that for most potentials $V(\phi)$, $P_s$ does exist globally up to a  critical value $s_c > 0$. 

\setcounter{equation}{0}
\section{Fake supergravity equations}
Consider now boost invariant, plane symmetric configurations,
\be
ds^2=r^2(-dt^2+dx_i dx^i)+{dr^2\over g(r)}, \qquad \phi=\phi(r),
\ee
 One finds that $g$ now satisfies an algebraic constraint
\be
g(r) =\frac{2r^2 V(\phi)}{r^2\phi'^2-d(d-1)},
\ee
and the scalar wave equation is
\be
\phi''+\left(\frac{g'}{2g}+\frac{d}{r} \right)\phi'-\frac{V_{,\phi}(\phi)}{g}=0.
\ee
We can reduce this second order equation to two first order ones via the ansatz
\be
\phi'(r)=-\frac{(d-1)P_{,\phi}(\phi)}{rP(\phi)},\qquad (d-1)P'(\phi)^2-dP(\phi)^2=V(\phi),
\ee
which implies
\be
g=\frac{2}{d-1}r^2P(\phi)^2.
\ee
Using the small $\phi$ expansion of $P$ in (\ref{genPs}), and plugging this into the fake supergravity equations we find
\be
\phi=\alpha \left[\frac{1}{r^{\Delta_-}}-\frac{s|\alpha|^{(\Delta_+-\Delta_-)/\Delta_-}}{r^{\Delta_+}}\right]+\ldots,
\ee
\be
g=r^2+\frac{\Delta_-\alpha^2}{(d-1)r^{2(\Delta_--1)}}-\frac{\Delta_+\Delta_-}{d(d-1)}\frac{4s|\alpha|^{d/\Delta_-}}{r^{d-2}}+\ldots
\ee
The energy density of such a solution is \cite{Hertog:2004ns}
\be
{E\over V}=(\Delta_+-\Delta_-)W + s\frac{\Delta_-(\Delta_+-\Delta_-)}{d}|\alpha|^{d/\Delta_-}.
\ee
Therefore the solution corresponding to $P_c$ saturates the bound analogous to (\ref{boundEd}) for planar solutions. As before, solutions corresponding to $s<s_c$ will have $P$ blow up exponentially and lead to naked singularities at $r=0$, and we exclude such solutions. 

\setcounter{equation}{0}
\section{Spherical solitons}

We now  relate the generalized minimum energy theorem in Appendix A to conjecture 1 in section 2.
If we consider static solutions with spherical symmetry, an appropriate ansatz is
\be
ds^2=-f(r)dt^2+\frac{dr^2}{g(r)}+r^2d\Omega_{k}, \qquad \phi=\phi(r),
\ee
where $d\Omega_k$ is the metric on a $(d-1)$-sphere with scalar curvature $R_{sphere}=(d-1)(d-2)k$. This parameterization allows us to easily study both asymptotically Poincar\`e and global AdS solutions. The equations of motion coming from (\ref{gendimaction}) are
\bea
\frac{d-1}{2}\left(\frac{g'}{g}-\frac{f'}{f}\right)+r\phi'^2=0,\nonumber
\eea
\bea
\frac{g'}{g}+\frac{f'}{f}+\frac{4rV(\phi)}{(d-1)g}+\frac{2(d-2)}{r}-\frac{2(d-2)k}{r g}=0,\nonumber
\eea
\be
\phi''+\left(\frac{g'}{2g}+\frac{f'}{2f}+\frac{d-1}{r}\right)\phi'-\frac{V'(\phi)}{g}=0.\nonumber
\ee
These equations have two symmetries. The first is simply rescaling  time:
$t\rightarrow \Lambda t, \ f\rightarrow f/\Lambda^2$.
There is also a scaling symmetry which changes the asymptotics by rescaling the size of the $S^{d-1}$,
\be
  \alpha = c \ti \alpha, \quad r = c \rho, \quad \phi(c\rho) = \ti \phi(\rho),  \quad g(c\rho)  = c^2 \ti g(\rho), \quad f(c\rho)  = c^2 \ti f(\rho)\label,\quad k=c^2\ti k.\label{genRescaling}
\ee
When we study the asymptotics, we find
\bea
\phi=\frac{\alpha}{r^{\Delta_-}}+\frac{\beta}{r^{\Delta_+}}+\ldots\nonumber,
\eea
\bea
g=r^2+k +\frac{\Delta_-\alpha^2}{(d-1)r^{2(\Delta_--1)}}-\frac{M_0}{r^{d-2}}+\ldots\nonumber
\eea
\be
f=r^2+k -\left(M_0+\frac{4\Delta_+\Delta_-\alpha\beta}{d(d-1)}\right)/r^{d-2}+\ldots
\ee
%To have a consistent classical theory we must impose boundary conditions on the scalar field. Ones that preserve our symmetries will be of the form
Given boundary conditions
\be
\beta=\beta(\alpha)=W'(\alpha),
\ee
for some function $W(\alpha)$, the total energy is \cite{Hertog:2004ns}
\be
E=\mathrm{Vol}(S^{d-1})\left[\frac{d-1}{2}M_0+\Delta_-\alpha\beta+(\Delta_+-\Delta_-)W\right].\label{gensolE}
\ee
If we consider static solitons (with $k\neq 0$), smoothness at the origin implies $g\approx k$, and an expansion at small $r$ gives us
\bea
\phi=\phi_0+\frac{V'(\phi_0)}{2 d k}r^2+\ldots\nonumber\eea
\bea
g=k-\frac{2V(\phi_0)}{d(d-1)}r^2+\ldots\nonumber\eea
\bea
f=k f_0-f_0 \frac{2V(\phi_0)}{d(d-1)}r^2+\ldots
\eea
Where $f_0$ is a free parameter corresponding to rescaling time, which we pick to obtain $f=k+r^2+\ldots$ at large radius. By changing $\phi_0$ we obtain a one parameter family of solitons, from which we define a function $\beta_0(\alpha)$ and therefore $W_0=-\int \beta_0 d\alpha$. 

We are interested in the behavior of $W_0(\alpha)$ for large $\alpha$. As in section 6, this can be extracted by the rescaling (\ref{genRescaling}). 
Under the technical assumption that the $c\rightarrow \infty$ limit is smooth, so that $\ti \phi$ can be expanded $\ti\phi=\ti\phi^{(0)}+\ti k\ti \phi^{(1)}+\ldots$, then an immediate generalization of the argument in section 6 shows that in the planar limit, $W_0$ takes the scale invariant form $\frac{s_w \Delta_-}{d}|\alpha|^{d/\Delta_-}$ for some $s_w$. It follows that we recover boost invariance, and we must have $s_w=s_c$. Therefore $W+W_0$ is bounded when $W+\frac{s_w \Delta_-}{d}|\alpha|^{d/\Delta_-}$ is bounded, and our minimum energy theorem proves conjecture 1.

\end{appendix}

% The bibliography goes here:
%%%%%%%%%%%%%%%%%%%%%%%%%%%%%%%%%%%%%%%%%%%%%%%%%%%%%%%%%%%%%%%%%%%%%%%

\end{document}